\documentclass{amsart}
\RequirePackage{amsmath,natbib,amsfonts}
\usepackage{graphicx}

\newcommand{\kr}{\ensuremath{Z}}

\newcommand{\genkr}{\hat{Z}}
\newcommand{\psbA}{\emph{psbA}}
\newcommand{\asinh}{\operatorname{asinh}}

\newcommand{\TABrounding}{
\begin{table}[ht]
\begin{center}
\begin{tabular}{llrrrrrr}
  \hline
  $S$ & $C$ & strict $Z_1$ & strict p & asinh $Z_1$ & asinh p & unit $Z_1$ & unit p \\
  \hline
  1 & 0.01 & 0.006578 & 0.0087 & 0.007016 & 0.0008 & 0.007054 & 0.0003 \\
  1 & 0.05 & 0.006584 & 0.0218 & 0.006986 & 0.0018 & 0.007036 & 0.0005 \\
  1 & 0.1 & 0.006562 & 0.035 & 0.007214 & 0.001 & 0.007322 & 0.0005 \\
  2 & 0.01 & 0.006601 & 0.0018 & 0.007076 & 0.0003 & 0.007281 & 0.0001 \\
  2 & 0.05 & 0.006587 & 0.0029 & 0.00696 & 0.0005 & 0.007111 & 0.0002 \\
  2 & 0.1 & 0.006592 & 0.0039 & 0.007088 & 0.0003 & 0.007423 &     0 \\
  3 & 0.01 & 0.006601 & 0.0017 & 0.006806 & 0.0005 & 0.006922 & 0.0002 \\
  3 & 0.05 & 0.006602 & 0.0018 & 0.006719 & 0.0003 & 0.006695 & 0.0001 \\
  3 & 0.1 & 0.006612 & 0.0012 & 0.006775 & 0.0003 & 0.006816 & 0.0001 \\
   \hline
\end{tabular}
\caption{Distances ($Z_1$) and significance levels (p) for various choices of clustering parameters and multiplicity interpretations described in the text for 10,000 randomizations.}
\label{tab:rounding}
\end{center}
\end{table}
}

\newcommand{\FIGcontrolTree}{
\begin{figure}[ht!]
\begin{center}
  \includegraphics[width=10cm]{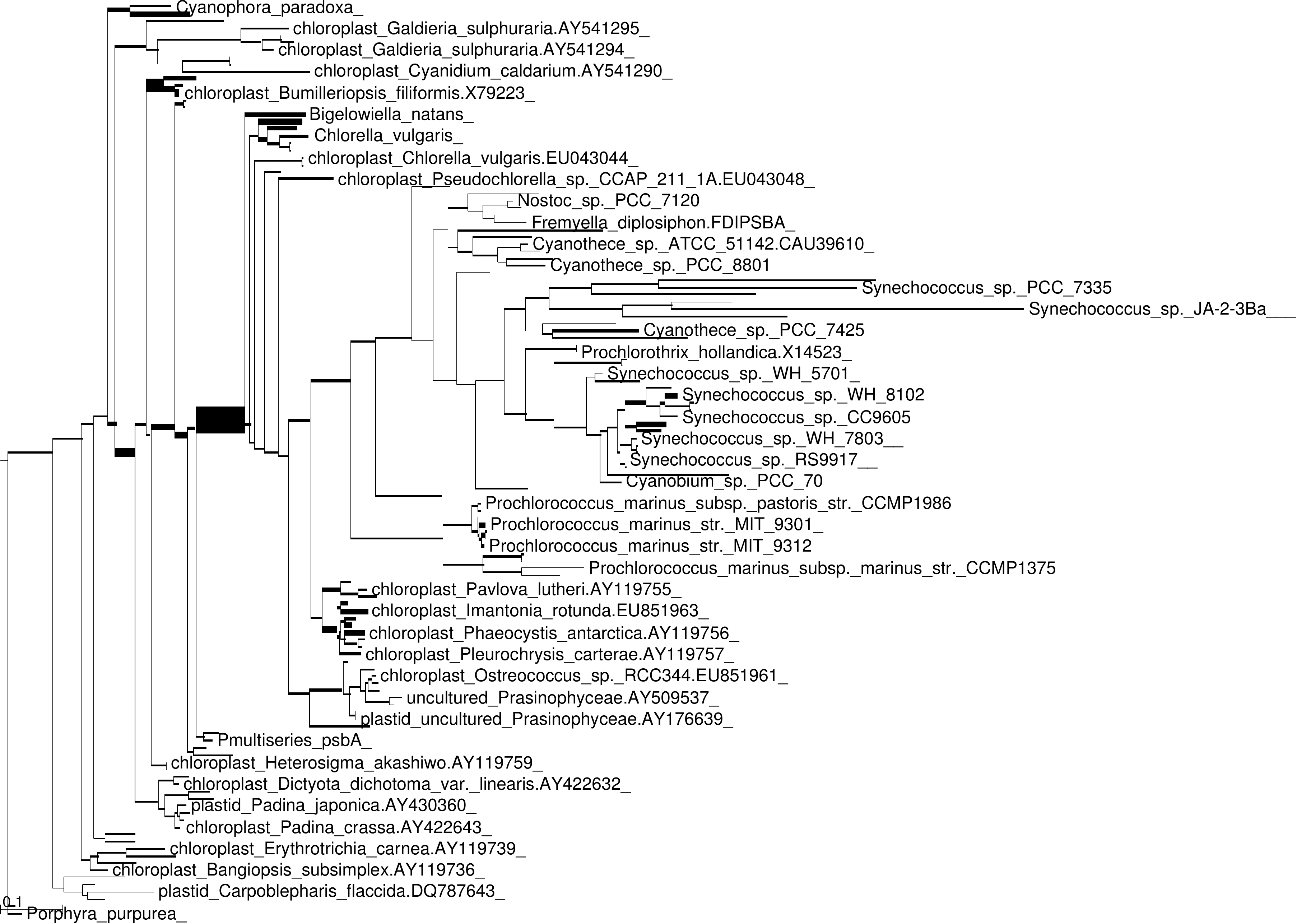}
\end{center}
\caption{Tree with branches thickened as a linear function of the number of placements in the control sample placed on that branch.}
\label{fig:controlTree}
\end{figure}
}

\newcommand{\FIGdmspTree}{
\begin{figure}
\begin{center}
  \includegraphics[width=10cm]{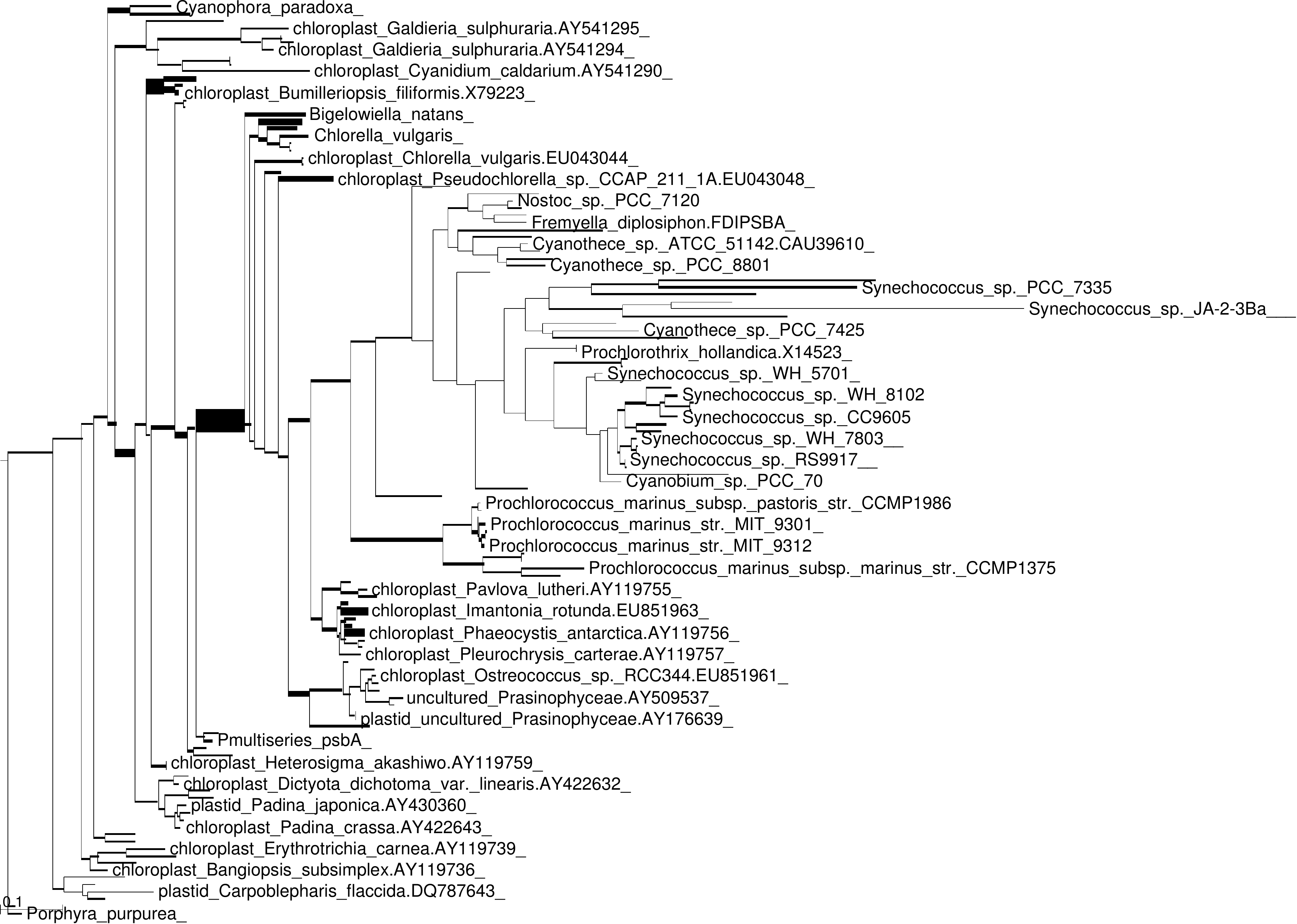}
\end{center}
\caption{Tree as in Figure~\ref{fig:controlTree} but for the DMSP-treated sample.}
\label{fig:dmspTree}
\end{figure}
}

\newcommand{\FIGshuffnorm}{
\begin{figure}
\begin{center}
  \includegraphics[width=8cm]{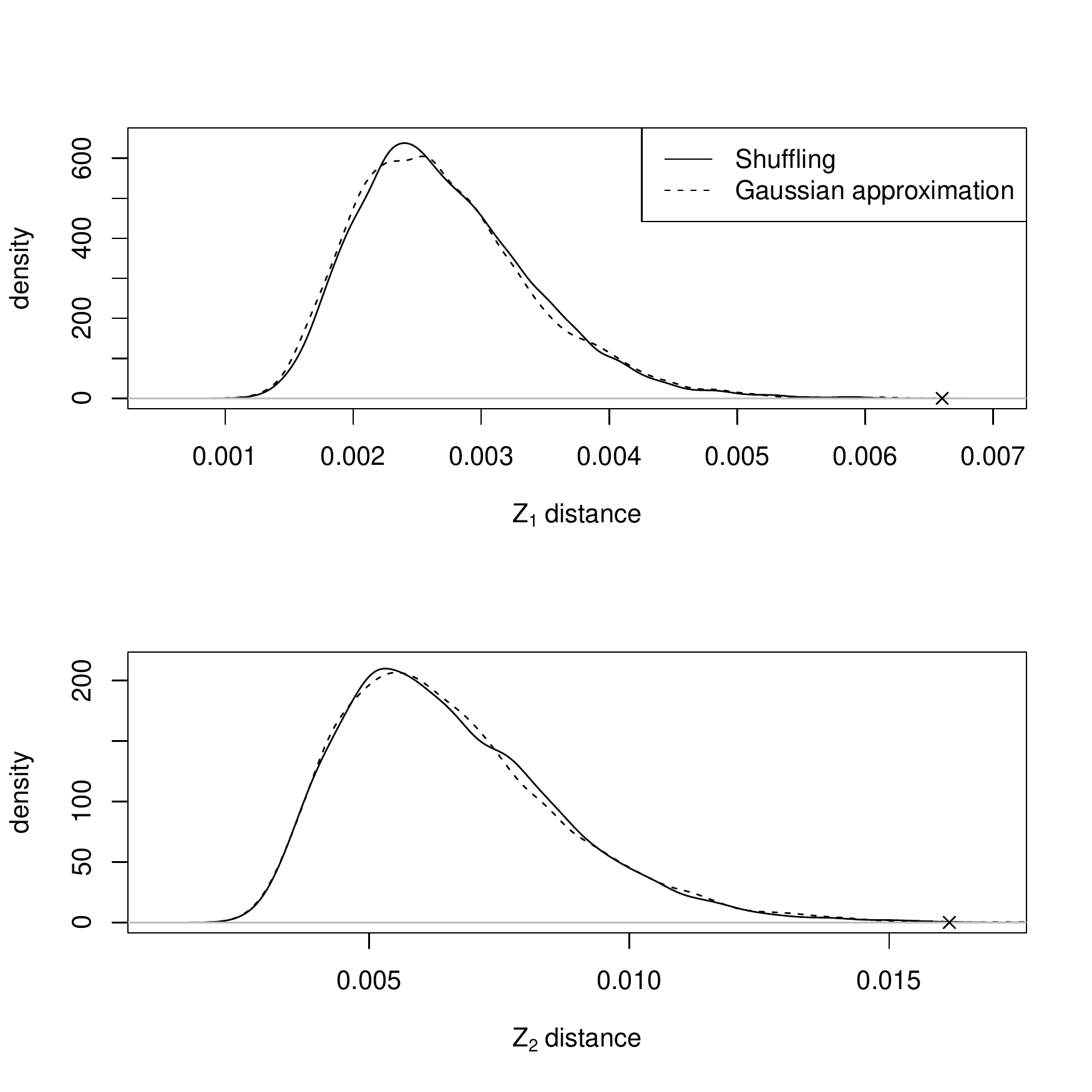}
\end{center}
\caption{Comparison of the distribution of $Z_1$ and $Z_2$ distances obtained by shuffling, Gaussian approximation, and the observed value (marked with x) for the example data set.}
\label{fig:shuffnorm}
\end{figure}
}

\newcommand{\FIGbary}{
\begin{figure}
\begin{center}
  \includegraphics[width=10cm]{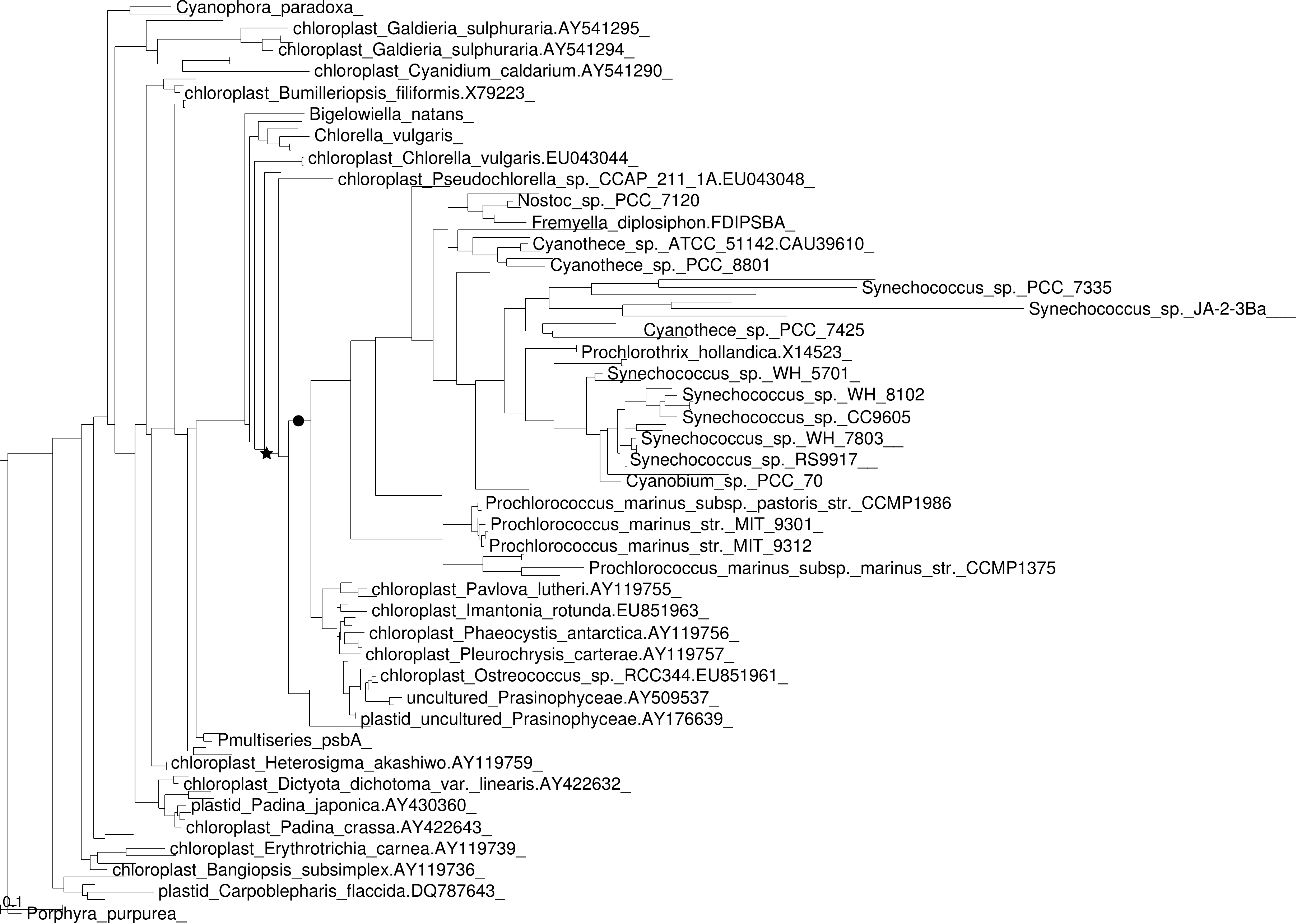}
\end{center}
\caption{Dendrogram with barycenters marked. Circle is the control sample, and star is the sample treated with DMSP.}
\label{fig:bary}
\end{figure}
}

\newcommand{\FIGbox}{
\begin{figure}
\begin{center}
  \includegraphics[width=8cm]{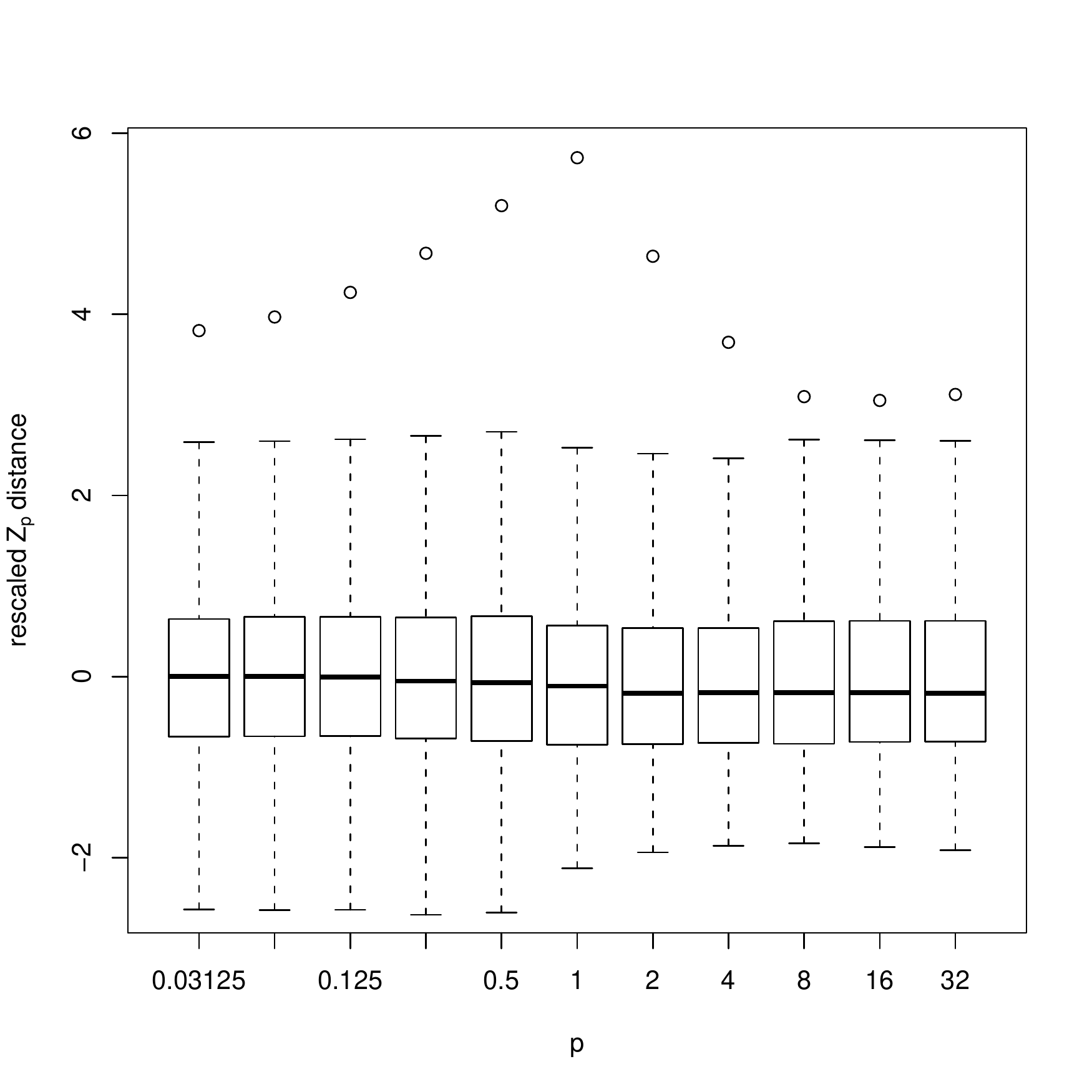}
\end{center}
\caption{Plot showing sample (point) and randomized ranges (box-and-whisker). Outliers eliminated for clarity. For each $p$, the distribution was rescaled by subtracting the mean and dividing by the standard deviation.
}
\label{fig:box}
\end{figure}
}

\newcommand{\FIGheat}{
\begin{figure}
\begin{center}
  \includegraphics[width=10cm]{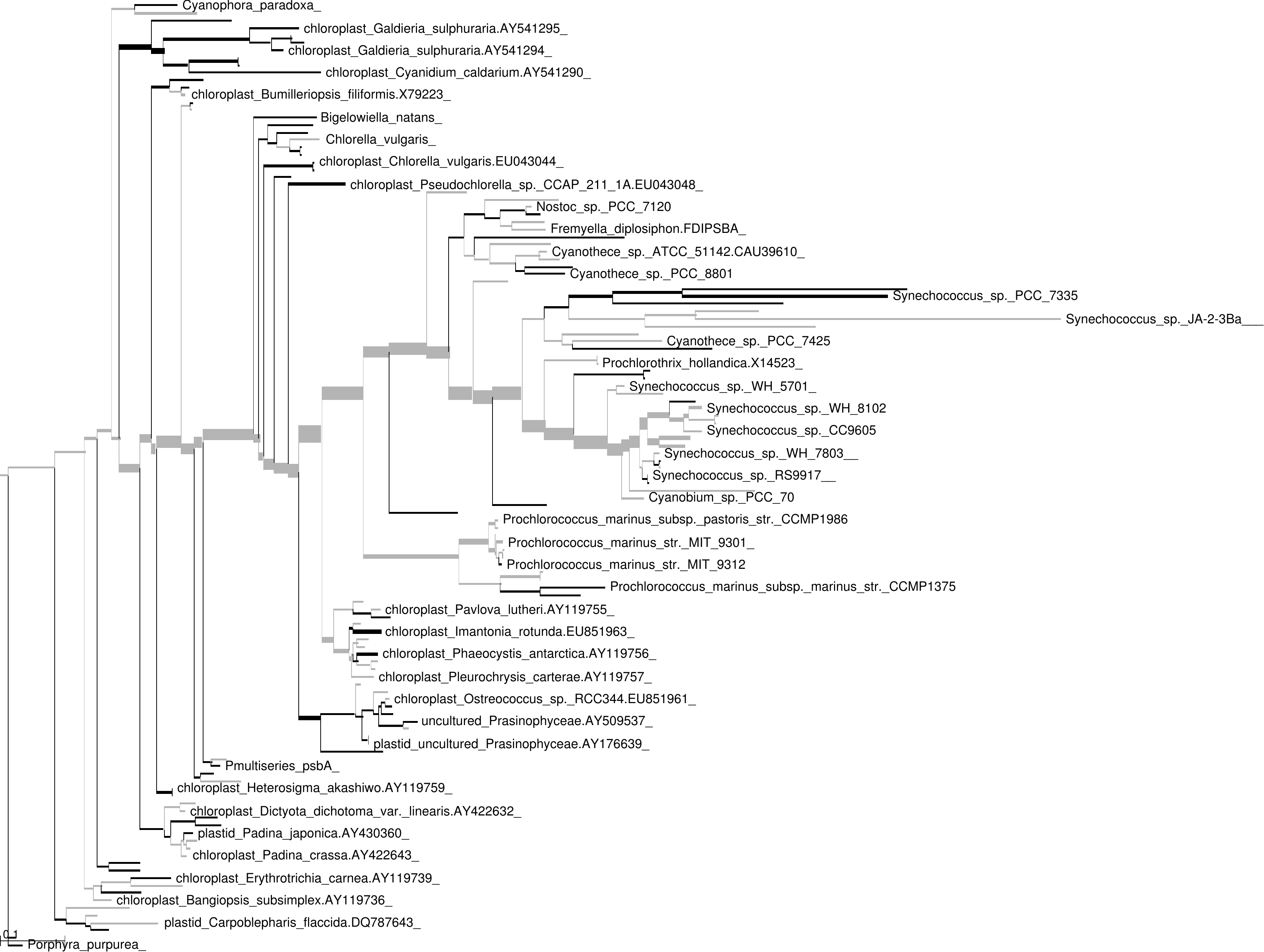}
\end{center}
\caption{
A tree displaying the optimal movement of mass for the KR metric.
When moving from the first probability distribution to the second, branches marked in gray have mass moving towards the root, while those marked in black have mass moving towards the leaves.
Thickness shows the quantity of mass moving through that branch.
}
\label{fig:heat}
\end{figure}
}

\newcommand{\FORarxiv}[1]{#1}
\newcommand{\FORsubmit}[1]{}

\newcommand{\eat}[1]{}

\newtheorem{remark}{Remark}[section]

\title[phylogenetic Kantorovich-Rubinstein metric]{The phylogenetic Kantorovich-Rubinstein metric for environmental sequence samples}
\author{Steven N. Evans}
\email{evans@stat.berkeley.edu}
\address{
Department of Statistics \#3860,
University of California,
367 Evans Hall,
Berkeley, CA 94720-3860
}

\author{Frederick A. Matsen}
\email{matsen@fhcrc.org}
\address{
Program in Computational Biology,
Fred Hutchinson Cancer Research Center,
1100 Fairview Ave. N. M1-B514,
P.O. Box 19024,
Seattle, WA 98109-1024
 }
\thanks{
  The first author was supported in part by NSF grant DMS-0907630.
  The second author was supported by the Miller Institute for Basic Research in Science, University of California at Berkeley, by FHCRC startup funds, and NIH grant HG005966-01.
}


\begin{document}

\FORsubmit{\maketitle}

\FORsubmit{\section{Summary}}
\FORarxiv{\begin{abstract}}
Using modern technology, it is now common to survey microbial communities by sequencing DNA or RNA extracted in bulk from a given environment.
Comparative methods are needed that indicate the extent to which two communities differ given data sets of this type.
{\em UniFrac}, a method built around a somewhat ad hoc phylogenetics-based distance between two communities, is one of the most commonly used tools for these analyses.
We provide a foundation for such methods by establishing that if one equates a metagenomic sample with its
empirical distribution on a reference phylogenetic tree,
then the {\em weighted UniFrac}  distance between two samples is just the classical
Kantorovich-Rubinstein (KR) distance between the corresponding empirical distributions.
We demonstrate that this KR distance and extensions of it that arise from incorporating
uncertainty in the location of sample points can be written as a
readily computable integral over the tree, we
develop $L^p$ Zolotarev-type generalizations of the metric, and we show how the p-value
of the resulting natural permutation test of the null hypothesis ``no difference between the two communities'' can be approximated using a functional of a Gaussian process indexed by the tree.
We relate the $L^2$ case to an ANOVA-type decomposition and find that the distribution of its associated Gaussian functional is that of a computable linear combination of independent $\chi_1^2$ random variables.
\FORarxiv{\end{abstract}}

\FORarxiv{\maketitle}

\section{Introduction}
\label{S:intro}

Next-generation sequencing technology enables sequencing of hundreds of thousands to millions of short DNA sequences in a single experiment.
This has led to a new methodology for characterizing the collection of microbes in a sample: rather than using observed morphology or the results of culturing experiments, it is possible to  sequence directly genetic material extracted in bulk from the sample.
This technology has revolutionized the possibilities for unbiased surveys of environmental microbial diversity, ranging from the human gut \citep{gill2006mah} to acid mine drainages \citep{baker2003mca}.
We consider statistical comparison procedures for such DNA samples.

We have divided this introductory section into several subsections.
We begin in Subsection~\ref{SS:unifrac_description}
by reviewing the definition of the {\em UniFrac} metrics
that were developed by microbial ecologists wishing to assign biologically meaningful distances between two samples of the type described above.
The metrics in the UniFrac papers are defined without preliminary justification via formulas.
Although it has been pointed out that alternative ways of using phylogenetic branch lengths are possible \citep{faith2009cladistic},
there has been little work investigating the extent to which there is a deeper mathematical rationale for these measures of similarity.
With the goal of building a more mathematically founded comparative framework, we next observe in Subsection~\ref{SS:samples_as_distributions} that DNA from an environmental sample for a given locus can be thought of naturally as a probability distribution on a reference phylogenetic tree, and proceed to propose  in Subsection~\ref{SS:comparing_distributions}
the Kantorovich-Rubinstein (KR) metric
as a familiar and sensible way of comparing two such probability distributions.
We then observe in the same subsection
how the KR metric can be computed via a simple integral over the tree,
and that the resulting distance is in fact a generalization of UniFrac.
The final subsection of the introduction, Subsection~\ref{SS:result_overview},
 summarizes the other results of the paper.

\subsection{Introduction to UniFrac and its variants}
\label{SS:unifrac_description}

In 2005, Lozupone and Knight introduced the UniFrac comparison measure to quantify the phylogenetic difference between microbial communities \citep{lozuponeKnightUniFrac05}, and in 2007 they and others proposed a corresponding {\em weighted} version \citep{lozuponeEaWeightedUniFrac07}.
These two papers already have hundreds of citations in total,
attesting to their centrality in microbial community analysis.
Researchers have used UniFrac to analyze microbial communities on the human hand \citep{fierer2008influence}, establish the existence of a distinct gut microbial community associated with inflammatory bowel disease \citep{frank2007molecular}, and demonstrate that host genetics play a major part in determining intestinal microbiota \citep{rawls2006reciprocal}.
The distance matrices derived from the UniFrac method are also
commonly employed as input to clustering algorithms, including hierarchical clustering
and UPGMA \citep{lozuponeEaWeightedUniFrac07}.
Furthermore, the distances are  widely used in conjunction with ordination methods such as PCA \citep{rintala2008diversity}
or to discover microbial community gradients with respect to another factor, such as ocean depth \citep{desnues2008biodiversity}.
Two of the major metagenomic analysis ``pipelines'' developed in 2010 had a UniFrac analysis as one of their endpoints  \citep{caporasoEaQIIME10, hartmanEaWATERS10}. Recently, the
software used to compute the two UniFrac distances has been re-optimized for speed \citep{hamadyEaFastUniFrac09} and it has been re-implemented in the heavily used {\em mothur} \citep{schlossEamothur09} microbial analysis software package.

The unweighted UniFrac distance uses only
presence-absence data and is defined as follows.
Imagine that one has two samples $A$ and $B$
of sequences.  Call each such sequence a {\em read}.
Build a phylogenetic tree on the total collection
of reads.
Color the tree according to the samples -- if a given branch sits on a path between two reads from sample $A$, then it is colored red, if it sits on a path between two reads from sample $B$,
then it is colored blue, and if both, then it is colored gray.
Unweighted UniFrac is then the fraction of the total branch length that is ``unique'' to one of the samples: that is, it is the fraction of the total branch length that is either red or blue.

Weighted UniFrac incorporates information about the frequencies of
reads from the two samples
by assigning weights to branch lengths that are not just $0$ or $1$.
Assume there are $m$ reads from sample $A$ and $n$ reads in sample $B$, and that one builds a phylogenetic tree $T$ from all $m+n$ reads.
For a given branch $i$ of the tree $T$,
let $\ell_i$ be the length of branch $i$ and define
$f_i$ to be the branch length fraction of branch $i$, i.e. $\ell_i$ divided by the total branch length of $T$.
The formula for the (raw) weighted UniFrac distance between the two samples is
\begin{equation}
  \label{eq:uniFrac}
\sum_{i=1}^n \ell_i \left| \frac{A_i}{m} - \frac{B_i}{n} \right|
\end{equation}
where $A_i$ and $B_i$ are the respective
number of descendants of branch $i$ from communities $A$ and $B$ \citep{lozuponeEaWeightedUniFrac07}.  In order to determine whether or
not a read is a descendant of a branch, one needs to prescribe
a vertex of the tree as being the root, but it turns out that different choices
of the root lead to the same value of the distance because
\begin{equation}
\label{eq:sand_transfer_special}
\left| \frac{A_i}{m} - \frac{B_i}{n} \right|
=
\frac{1}{2}
\left(
\left|
\frac{A_i}{m} - \frac{B_i}{n}
\right|
+
\left|
\left( 1 - \frac{A_i}{m} \right)
- \left( 1 - \frac{B_i}{n} \right)
\right|
\right),
\end{equation}
and the quantity on the right only depends on the proportions of reads in
each sample that are in the two disjoint subtrees obtained by deleting the
branch $i$.  Also, similar reasoning shows that the (unweighted)
UniFrac distance is, up to a factor of $\frac{1}{2}$,
 given by a formula similar to
\eqref{eq:uniFrac} in which $A_i/m$ (respectively, $B_i/n$)
is replaced by a quantity that is either $1$ or $0$ depending on whether
there are any descendants of branch $i$ in the $A$ (respectively, $B$) sample and
the branch length $\ell_i$ is replaced by the branch length fraction $f_i$.
Using the quantities $\ell_i$ rather than the
$f_i$ simply changes the resulting distance by a multiplicative constant,
the total branch length of the tree $T$.

The UniFrac distances can also be calculated using a pre-existing tree
(rather than one built from samples)
by performing a sequence comparison such as BLAST to associate
a read with a previously identified sequence and attaching
the read to that sequence's leaf in the pre-existing tree
with an intervening branch of zero length.
Using this mapping strategy, the tree used for comparison can be adjusted depending on the purpose of the analysis.
For example, the user may prefer an ``ultrametric'' tree (one with the same total branch length from the root to each tip) instead of one made with branch lengths that reflect amounts of molecular evolution.

With the goal of making reported UniFrac values comparable
across different trees, it is common to divide by a suitable
scalar to fit them into the unit interval.
Given a rooted tree $T$ and counts $A_i$ and $B_i$ as above, the raw weighted UniFrac value is bounded above by
\begin{equation}
  \label{eq:uniFracD}
  D = \sum_{i=1}^n d_i \left( \frac{A_i}{m} + \frac{B_i}{n} \right)
\end{equation}
where $d_i$ is the distance from the root to the leaf side of edge $i$ \citep{lozuponeEaWeightedUniFrac07}.
When divided by this factor, the resulting scaled UniFrac values sit in the unit interval; a scaled UniFrac value of one means that there exists a branch adjacent to the root which can be cut to separate the two samples.
Note that the factor $D$, and consequently the ``normalized'' weighted UniFrac value, does depend on the position of the root.

A statistical significance for the observed UniFrac distance
is typically assigned by a permutation procedure
that we review here for the sake of completeness.
The idea of a permutation test (also known as a randomization test) goes back
to \citet{Fisher35} and \citet{Pitman37a, Pitman37b, Pitman38} (see Good, 2005, and Edgington and Onghena, 2007,
\nocite{MR2103758, MR2291573}
for guides to the more recent literature).
Suppose that our data are a pair of samples with counts $m$ and $n$, respectively, and
that we have computed the UniFrac distance between the samples.
Imagine creating a new pair of ``samples'' by taking some other subset of size
$m$ and its complement from the set of all $m+n$ reads
and then computing the distance between the
two new samples.
The proportion of the $\binom{m+n}{m}$ choices of such pairs of samples that result in a distance larger
than the one observed in the data is an indication of the significance of the observed distance.
Of course,
we can rephrase this procedure as taking a uniform random subset of reads of size $m$
and its complement
(call such an object a {\em random pair of pseudo-samples})
and asking for the probability that the distance between these
 is greater than the observed one.
Consequently, it is possible
(and computationally necessary for large values of $m$ and $n$)
to approximate the proportion/probability in question
by taking repeated independent choices of the random
subset and recording the proportion of
choices for which there is a distance
between the pair of pseudo-samples greater than the observed one.
We call the distribution of the distance between a random pair of pseudo-samples
produced from a uniform random subset of reads of size $m$ and its complement of size $n$ the
{\em distribution of the distance under the null hypothesis of no clustering}.

\subsection{Phylogenetic placement and probability distributions on a phylogenetic tree}
\label{SS:samples_as_distributions}

We now describe how it is natural to begin with
a fixed \emph{reference phylogenetic tree} constructed from previously-characterized
DNA sequences and then use likelihood-based phylogenetic methods
to map a DNA sample from some environment to a collection of \emph{phylogenetic placements} on the reference tree.
This collection of placements can then be thought of as a probability distribution on the reference tree.

In classical likelihood-based phylogenetics (see, e.g., Felsenstein, 2004), \nocite{felsensteinBook04}
one has data consisting of
DNA sequences from a collection of {\em taxa} (e.g. species)
and a probability model for that data.
The probability model is composed of two ingredients.
The first ingredient is a
tree with branch lengths that has its
leaves labeled by the taxa and describes their evolutionary relationship.  The second ingredient
is a Markovian stochastic mechanism for the evolution of DNA
along the branches of the tree.  The parameters of the model are the tree
(its topology and branch lengths) and the rate parameters in the DNA
evolution model.  The likelihood of the data is, as usual, the function
on the parameter space that gives the probability of the observed data.
The tree and rate parameters can be estimated using standard approaches
such as maximum likelihood or Bayesian methods.

Suppose one already has, from whatever source, DNA sequences for each of a number of taxa
along with a corresponding phylogenetic tree and rate parameters, and that a new
sequence, the {\em query sequence}, arrives.  Rather than estimate
a new tree and rate parameters {\em ab initio}, one can take the rate parameters as given
and only consider trees that consist of the existing tree, the
reference tree, augmented by a branch of some length leading from an
attachment point on the reference tree to a leaf labeled by the new taxon.
The relevant likelihood is now the conditional probability of the query sequence
as a function of the attachment point and the pendant branch length, and one
can input this likelihood into maximum likelihood or Bayesian methods to
estimate these two parameters.
For example,
a maximum-likelihood {\em point phylogenetic placement} for a given query sequence is the maximum-likelihood estimate of the attachment
point of the sequence to the tree and the pendant branch length leading to the sequence.
Such estimates are produced by a number of algorithms
 \citep{vonMeringEaQuantitative08, monierEaLargeViruses08, bergerStamatakisEPA09, matsenEaPplacer10}.
Typically, if there is more than one
query sequence, then this procedure is applied in isolation to each one using
the same reference tree; that is, the taxa corresponding to the successive
query sequences aren't used to enlarge the reference tree.
By fixing a reference tree rather than attempting to build a phylogenetic tree
for the sample {\em de novo}, recent algorithms of this type are able to place tens of thousands of query sequences per hour per processor on a reference tree of one thousand taxa, with linear performance scaling in the number of reference taxa.

For the purposes of this paper, the data we retain from a collection of point phylogenetic placements will simply be the attachment locations of those placements on the reference phylogenetic tree. We will call these positions
{\em placement locations}.
We can identify such a set of placement locations with its empirical distribution, that is, with the probability distribution that places an equal mass at each placement.
In this way, starting with a reference tree and an aligned collection of reads, we arrive at a probability distribution on the reference tree representing the distribution of those reads across the tree.

One can also adopt a Bayesian perspective and assume a prior probability on the branch to which the attachment is made,
the attachment location within that branch,
and the pendant branch length, in order to calculate
a posterior probability distribution for a placement.
For example, one might take a prior for the attachment location and pendant branch length
that assumes these quantities are independent, with the prior distribution for the
attachment location being uniform over branches and uniform within each branch
and with the prior distribution of the pendant branch length
being exponential or uniform over some range.
By integrating out the pendant branch length, one obtains a posterior probability distribution $\mu_i$ on the tree for query sequence $i$.  We call such a probability distribution
a {\em spread placement}: with priors such as those
above, $\mu_i$ will have a density with respect
to the natural length measure on the tree.
It is natural to associate this collection
of probability distributions with the single distribution $\sum_i \mu_i / n$, where $n$
is the number of query sequences.

For large data sets, it is not practical to record detailed information about the posterior probability distribution.
Thus, in the implementation of \citet{matsenEaPplacer10}, the posterior probability is computed on a branch-by-branch basis for a given query sequence by integrating out the attachment location and the pendant branch length, resulting in a probability for each branch.
The mass is then assigned to the attachment location of the maximum likelihood phylogenetic placement.
With this simplification, we are back in the point placement situation in which each query sequence
is assigned to a single point on the reference tree and the collection of assignments is
described by the empirical distribution of this set of points.
However, since it is possible in principle to work with a representation
of a sample that is not just a discrete distribution with
equal masses on each point, we develop the theory
in this greater level of generality.

\subsection{Comparing probability distributions on a phylogenetic tree}
\label{SS:comparing_distributions}

If one wished to use the standard Neyman-Pearson
framework for statistical inference to determine
 whether two metagenomic samples
came from communities with the ``same'' or ``different'' constituents,
one would first have to
propose a family of probability distributions that described the outcomes of sampling
from a range of communities and then construct a test of the hypothesis that the two samples were realizations
from the same distribution in the family.
However, there does not appear to be such a family of distributions that is appropriate in this setting.

We are thus led to the idea of
representing the two samples as probability distributions on the reference tree in the manner described in
Subsection~\ref{SS:samples_as_distributions} above,
calculating a suitable distance between these two probability distributions,
and using the general permutation/randomization test approach
reviewed in Subsection~\ref{SS:unifrac_description} above
to assign a statistical significance to the observed distance.

The key element in implementing this
proposal is the choice of a suitable metric on the
space of probability distributions on the reference tree.
There are, of course, a multitude of choices:
Chapter 6 of \citet{MR2459454} notes that there are ``dozens and dozens'' of them
and provides a discussion of their similarities, differences and various virtues.

Perhaps the most familiar metric is the total variation distance, which is just the supremum over all (Borel)
sets of the difference between the masses assigned to the set by the two distributions.
The total variation distance is clearly inappropriate for our purposes, however, because it
pays no attention to the evolutionary distance structure on the tree: if one took $k$ point
placements and constructed another set of placements by perturbing each of the original placements
by a tiny amount so that the two sets of placements were disjoint, then the total variation
distance between the corresponding probability distributions would be $1$,
the largest it
can be for any pair of probability distributions, even though we would regard the two sets of
placements as being very close. Note that even genetic material from organisms of the same species can
result in slightly different placements due to genetic variation within species and experimental error.

We therefore need a metric that is compatible with the evolutionary distance on
the reference tree and measures two distributions as being close if one is obtained
from the other by short range redistributions of mass.   The Kantorovich-Rubinstein (KR) metric,
which can be defined for probability distributions on an arbitrary metric space,
is a classical and widely used distance that meets this requirement and, as we shall see,
has other desirable properties such as being easily computable on a tree.
It is defined rigorously in Section~\ref{sec:kr} below, but it can be described intuitively in physical terms as follows.
Picture each of two probability distributions on a metric space as a
collection of piles of sand with unit total mass: the mass of sand in the pile at a given point is equal to the probability mass at that point.
Suppose that the amount of ``work'' required to transport an amount of sand from one place to another is proportional to the mass of the sand moved times the distance it has to travel.
Then, the KR distance between two probability distributions $P$ and $Q$ is simply the minimum amount of work required to move sand in the configuration corresponding to $P$ into the configuration corresponding to $Q$.
It will require little effort to move sand between the configurations corresponding to two similar probability distributions, while more will be needed for two distributions that place most of their respective masses on disjoint regions of the metric space.  As noted by \citet{MR2459454}, the KR metric is also
called the {\em Wasserstein(1) metric} or, in the engineering literature, the {\em earth mover's distance}.
We note that mass-transport ideas have already been used in evolutionary bioinformatics for the comparison and clustering of ``evolutionary fingerprints''-- such a fingerprint being defined by \citet{kosakovsky2010evolutionary} as a discrete bivariate distribution on synonymous and nonsynonymous mutation rates for a given gene.

\subsection{Overview of results}
\label{SS:result_overview}

Our first result is that in the phylogenetic case, the optimization implicit in the definition of the KR metric can be done analytically, resulting in a closed form expression that can be evaluated in linear time, thereby enabling analysis of the volume of data produced by large-scale sequencing studies.
Indeed, as shown in Section~\ref{sec:kr}, the metric can be represented as a single integral over the tree, and for point placements the integral reduces to a summation with a number of terms on the order of the number of placements.
In contrast, computing the KR metric in Euclidean spaces of dimension greater than one requires a linear programming optimization step.
It is remarkable that the point version of this closed-form expression for the phylogenetic KR distance
(although apparently not the optimal mass transport justification for the distance)
was intuited by microbial ecologists and is nothing
other than the {\em weighted UniFrac} distance recalled in Subsection~\ref{SS:unifrac_description} above.

We introduce  $L^p$ generalizations of the KR metric
that are analogous to ones on the real line due to Zolotarev \citep{MR1105086, MR1619170} -- the KR metric corresponds to the case $p=1$.
Small $p$ emphasizes primarily differences due to separation of samples across the tree, while large $p$ emphasizes large mass differences.
The generalizations do not arise from optimal mass transport considerations, but we
remark in Section~\ref{sec:ANOVA} that the square of
the $p=2$ version does have an appealing ANOVA-like interpretation
as the amount of variability in a pooling of the two samples that is not accounted
for by the variability in each of them.

We show in Section~\ref{sec:significance} that
the distribution of the distance under the null hypothesis of no clustering is
approximately that of a readily-computable functional of a Gaussian process indexed by
the tree and that this Gaussian process is relatively simple to simulate.  Moreover, we observe that when
$p=2$ this approximate distribution is that of the square root of a weighted sum of $\chi_1^2$
random variables.
We also discuss the interpretation of the resulting $p$ value when
the data exhibit local ``clumps'' that might be viewed as being
the objects of fundamental biological
interest rather than the individual reads.

In Section~\ref{sec:discussion}, we discuss alternate approaches to sample comparison.  In particular, we remark that any probability distribution on a tree has a well-defined
barycenter (that is, center of mass) that can
be computed effectively.   Thus, one can obtain a
one point summary of the location of a sample by considering
the barycenter of the associated probability distribution
and measure the similarity of two
samples by computing the distance between the corresponding
barycenters.

\section{The phylogenetic Kantorovich-Rubinstein metric}
\label{sec:kr}

In this section we more formally describe
the phylogenetic Kantorovich-Rubinstein metric, which is a particular case
of the family of Wasserstein metrics.
We then use a dual formulation of the KR metric to show that it
can be calculated in linear time via a simple integral over the tree.
We also introduce a Zolotarev-type $L^p$ generalization.

%\subsection{Closed formula and properties}
%Having a single subsection after a brief introduction looked strange.

Let $T$ be a tree with branch lengths.
Write $d$ for the path distance on $T$.
We assume that probability distributions have been given on the tree via collections of either ``point'' or ``spread'' placements as described in the introduction.

For a metric space $(S,r)$, the Kantorovich-Rubinstein distance $\kr(P,Q)$ between two Borel probability distributions $P$ and $Q$ on $S$ is defined as follows.
Let $\mathcal{R}(P,Q)$ denote the set of probability distributions $R$ on the product space $S \times S$ with the property $R(A \times S) = P(A)$ and $R(S \times B) = Q(B)$ for all Borel sets $A$ and $B$ (that is, the two marginal distributions of $R$ are $P$ and $Q$).
Then,
\begin{equation}
\label{eq:kr_original}
\kr(P,Q) : = \inf\left\{ \int_{S \times S} r(x,y) \, R(dx,dy) : R \in \mathcal{R}(P,Q) \right\};
\end{equation}
see, for example, \citep{MR1105086, MR1619170, MR1964483, MR2401600, MR2459454}.

There is an alternative formula for $\kr(P,Q)$ that comes from convex duality.
Write $\mathcal{L}$ for the set of functions $f : S \rightarrow \mathbb{R}$ with the Lipschitz property
$|f(x) - f(y)| \le r(x,y)$ for all $x,y \in S$.
Then,
\[
\kr(P,Q) = \sup\left\{ \int_S f(x) \, P(dx) - \int_S f(y) \, Q(dy) : f \in \mathcal{L} \right\}.
\]

We can use this expression to get a simple explicit formula for $\kr(P,Q)$ when $(S,r) = (T,d)$.

Given any two points $x,y \in T$, let $[x,y]$ be the arc between them.
There is a unique Borel measure $\lambda$ on $T$ such that $\lambda([x,y]) = d(x,y)$ for all $x,y \in T$.
We call $\lambda$ the {\em length measure};
it is analogous to Lebesgue measure on the real line.
Fix a distinguished point $\rho \in T$, which we call the ``root'' of the tree.
For any $f \in \mathcal{L}$ with $f(\rho) = 0$, there is an $\lambda$-a.e. unique Borel function $g:T \rightarrow [-1,1]$ such that $f(x) = \int_{[\rho,x]} g(y) \, \lambda(dy)$ (this follows easily from the analogous fact for the real line).

Given $x \in T$, put $\tau(x) := \{y \in T : x \in [\rho,y]\}$; that is, if we draw the tree
with the root $\rho$ at the top of the page, then $\tau(x)$ is the sub-tree below $x$.
Observe that if $h: T \rightarrow \mathbb{R}$ is a bounded Borel function and $\mu$ is a Borel probability distribution on $T$, then we have the integration-by-parts formula
\[
\begin{split}
\int_T \left( \int_{[\rho,x]} h(y) \, \lambda(dy) \right) \, \mu(dx)
& =
\int_{T \times T} 1_{[\rho,x]}(y) h(y) \, (\mu \otimes \lambda) (dx,dy) \\
& =
\int_{T \times T} 1_{\tau(y)}(x) h(y) \, (\mu \otimes \lambda) (dx,dy) \\
& =
\int_T h(y) \left( \int_{\tau(y)} \, \mu(dx) \right) \, \lambda(dy) \\
& =
\int_T h(y) \mu(\tau(y)) \, \lambda(dy). \\
\end{split}
\]

Thus, if $P$ and $Q$ are two Borel probability distributions on $T$
and $f:T \to \mathbb{R}$ is given by $f(x) = \int_{[\rho,x]} g(y) \, \lambda(dy)$, then we have
\[
  \int_T f(x) \, P(dx)
  = \int_T \left( \int_{[\rho, x]} g(y) \lambda(dy) \right) \, P(dx)
  = \int_T g(y) P(\tau(y)) \, \lambda(dy),
\]
and an analogous formula holds for $Q$.  Hence,
\[
\kr(P,Q)
=
\sup
\left\{
\int_T g(y) \left[ P(\tau(y)) - Q(\tau(y)) \right] \, \lambda(dy) : -1 \le g \le +1
\right\}.
\]
It is clear that the integral is maximized by taking $g(y) = +1$ (resp. $g(y) = -1$) when
$P(\tau(y)) > Q(\tau(y))$ (resp. $P(\tau(y)) < Q(\tau(y))$), so that
\begin{equation}
  \label{eq:krInt}
\kr(P,Q)
=
\int_T  \left| P(\tau(y)) - Q(\tau(y)) \right| \, \lambda(dy).
\end{equation}

Note that \eqref{eq:uniFrac} is the special case of
\eqref{eq:krInt} that arises
when $P$ assigns point mass $1/m$ to each of the leaves in community $A$, and $Q$ assigns point mass $1/n$ to each of the leaves in community $B$.

We can generalize the definition of the
Kantorovich-Rubinstein distance by taking any pseudo-metric
$f$ on $[0,1]$ and setting
\[
\genkr_f(P,Q)
=
\int_T  f(P(\tau(y)),Q(\tau(y))) \, \lambda(dy).
\]
This object will be a pseudo-metric on the
space of probability distributions on the tree $T$.
All of the distances considered so far are of the
form $\genkr_f$ for an appropriate choice of the pseudo-metric $f$:
(unweighted) UniFrac results from taking $f(x,y)$ equal to one when
exactly one of $x$ or $y$ is greater than zero,
and $\kr$ arises when $f(x,y)=|x-y|$.

Furthermore, if $f(x,y) = f(1-x, 1-y)$, then $\genkr_f$ is invariant with
respect to the position of the root.
Indeed, for $\lambda$-a.e. $y \in T$
we have that $y$ is in the interior
of a branch and  $P(\{y\}) = Q(\{y\}) = 0$ so that, for such $y$,
$P(\tau(y))$ and $P(T \setminus \tau(y))$
(respectively, $Q(\tau(y))$ and $Q(T \setminus \tau(y))$)
are the $P$-masses (respectively, $Q$-masses) of the
two disjoint connected components of $T$ produced by removing $y$
(cf. \eqref{eq:sand_transfer_special}), and hence these
quantities don't depend on the choice of the root.
Because
\begin{equation*}
\begin{split}
f(P(\tau(y)), Q(\tau(y)))
& =
\frac{1}{2}
\left[
f(P(\tau(y)), Q(\tau(y))) + f(1 - P(\tau(y)), 1- Q(\tau(y)))
\right] \\
& =
\frac{1}{2}
\left[
f(P(\tau(y)), Q(\tau(y)))
+
f(P(T \setminus \tau(y)), Q(T \setminus \tau(y)))
\right]
\end{split}
\end{equation*}
for any $y \in T$,
the claimed invariance follows upon integrating with
respect to $\lambda$.
In particular, we see that the distance
$\kr$ is invariant to the position of the root,
a fact that is already apparent from the original definition
\eqref{eq:kr_original}.

In a similar spirit, the KR distance as defined by
the integral \eqref{eq:krInt} can be generalized
to an $L^p$ Zolotarev-type version by setting
\[
\kr_p(P,Q) =
\left[\int_T \left| P(\tau(y)) - Q(\tau(y)) \right|^p \, \lambda(dy)\right]^{\frac{1}{p} \wedge 1}
\]
for $0 < p < \infty$
-- see \citep{MR1105086, MR1619170} for a discussion of analogous metrics for probability distributions
on the real line.
Intuitively, large $p$ gives more weight in the distance to parts of the tree which are maximally different in terms of $P$ and $Q$, while small $p$ gives more weight to differences which require lots of transport.
The position of the root $\rho$ also does not matter for this generalization
of $\kr$ by the argument above.

As the following computations show, the distance $\kr_2$ has
a particularly appealing interpretation.  First note that
\[
\begin{split}
\kr_2^2(P,Q) & = \int_T |P(\tau(u)) - Q(\tau(u))|^2 \, \lambda(du) \\
& \quad =
\int_T P(\tau(u))^2 \, \lambda(du)
- 2 \int_T P(\tau(u)) Q(\tau(u)) \, \lambda(du) \\
& \qquad + \int_T Q(\tau(u))^2 \, \lambda(du) \\
& \quad =
\int_T \left[\int_T \int_T 1_{[\rho,v]}(u) 1_{[\rho,w]}(u) \, P(dv) \, P(dw) \right] \, \lambda(du) \\
& \qquad - 2 \int_T \left[\int_T \int_T 1_{[\rho,v]}(u) 1_{[\rho,w]}(u) \, P(dv) \, Q(dw) \right] \, \lambda(du) \\
& \qquad + \int_T \left[\int_T \int_T 1_{[\rho,v]}(u) 1_{[\rho,w]}(u) \, Q(dv) \, Q(dw) \right]  \, \lambda(du) \\
\end{split}
\]
Now, the product of indicator functions $1_{[\rho,v]} 1_{[\rho,w]}$
is the indicator function of the set $[\rho,v] \cap [\rho,w]$.  This
set is an arc of the form $[\rho, v \wedge w]$, where
$v \wedge w$ is the ``most recent common ancestor'' of $v$ and $w$ relative to the root $\rho$.
Hence, $\int_T  1_{[\rho,v]}(u) 1_{[\rho,w]}(u)  \, \lambda(du) = \lambda([\rho, v \wedge w])$
is $d(\rho, v \wedge w) = \frac{1}{2} \left[d(\rho,v) + d(\rho,w) - d(v,w)\right]$.
Therefore,
\begin{equation*}
\begin{split}
\kr_2^2(P,Q) & =
\frac{1}{2}
\biggl[
2 \int_T \int_T d(v,w) \, P(dv) \, Q(dw) \\
& \quad - \int_T \int_T d(v,w) \, P(dv) \, P(dw)
- \int_T \int_T d(v,w) \, Q(dv) \, Q(dw)
\biggr].\\
\end{split}
\end{equation*}
Thus, if $X', X'', Y', Y''$ are independent $T$-valued random variables, where
$X', X''$ both have distribution $P$ and $Y', Y''$ both have distribution $Q$,
then
\begin{equation}
\label{eq:Z2_equivalent}
\kr_2^2(P,Q) = \frac{1}{2} \left(\mathbb{E}[d(X',Y') - d(X',X'')] +
\mathbb{E}[d(X',Y') - d(Y',Y'')]\right).
\end{equation}

Analogous to weighted UniFrac, one can ``normalize'' the KR distance by dividing it by a scalar.
The most direct analog of the scaling factor $D$
used for weighted UniFrac on a rooted tree \eqref{eq:uniFracD} would be twice the KR
distance between $(P+Q)/2$ and a point mass located at the root.
This is an upper bound by the triangle inequality.
A root-invariant version would be to instead place the point mass at the
center of mass (that is, the barycenter,
see Section~\ref{sec:barycenter})
of $(P+Q)/2$, and twice the analogous distance is again an upper bound by the
triangle inequality. It is clear from the original definition of the
KR distance \eqref{eq:kr_original}, that
$Z_1(P,Q)$ is bounded above by the diameter
of the tree (that is, $\max_{x,y} d(x,y)$) or
by the possibly smaller similar quantity that arises by restricting
$x$ and $y$ to the respective supports of $P$ and $Q$.
Any of these upper bounds can be used as a ``normalization
factor''.

The goal of introducing such normalizations would be to
permit better comparisons between distances obtained for different
pairs of samples.  However, some care needs to be exercised here:
it is not clear how to scale distances for pairs on two
very different reference trees so that similar values
of the scaled distances convey any readily interpretable
indication of the extent to which the elements of the
two pairs differ from each other in a ``similar'' way.
In short, when comparing results between trees, the
KR distance and its generalizations are
more useful as test statistics than as
descriptive summary statistics.

\section{Assessing significance}
\label{sec:significance}

To assess the significance of the observed distance between
the probability distributions associated with a pair of
 samples of placed reads
of size $m$ and $n$, we use the permutation
strategy mentioned in the Introduction for
assigning significance to observed UniFrac distances.  In general,
we have a pair of probability distributions representing
the pair of samples that is of the form
$P = \frac{1}{m}\sum_{i=1}^m \pi_i$
and
$Q = \frac{1}{n}\sum_{j=m+1}^n \pi_j$,
where $\pi_k$ is a probability distribution on the reference
tree $T$ representing the placement of the $k^{\mathrm{th}}$ read
in a pooling of the two samples (in the point placement
case, each $\pi_k$ is just a unit point mass at some point
$w_k \in T$).
We imagine creating all $\binom{m+n}{m}$ pairs
of ``samples'' that arise from placing $m$ of the reads
from the pool into one sample and the
remaining $n$ into the other, computing the distances between the two
probability distributions on the reference tree that result from the placed
reads, and determining what proportion of these distances exceed the distance
observed in the data.  This proportion may be thought of as a p-value for a
test of the null hypothesis of no clustering against an alternative of some
degree of clustering.

Of course, for most values of $m$ and $n$
it is infeasible to actually perform this exhaustive
listing of distances.
We observe that if
$I \subseteq \{1, \ldots, m+n\}$ is a uniformly distributed random subset with cardinality $m$ (that is, all $\binom{m+n}{m}$ values are equally likely),
$J := I^c$ is the complement of $I$, $\tilde P$ is the random probability
distribution $\frac{1}{m}\sum_{i \in I} \pi_i$, and
$\tilde Q$ is the random probability distribution $\frac{1}{n}\sum_{j \in J} \pi_j$,
then the proportion of interest is simply the probability that
the distance between $\tilde P$ and $\tilde Q$ exceeds the distance between
$P$ and $Q$.  We can approximate this probability in the obvious way by taking
independent replicates of $(I,J)$ and hence of $(\tilde P, \tilde Q)$ and looking
at the proportion of them that result in distances
greater than the observed one.  We illustrate this Monte Carlo
approximation procedure in Section~\ref{sec:application}.

\subsection{Gaussian approximation}
\label{sec:gaussian}

Although the above Monte Carlo approach to approximating a p-value
is conceptually straightforward,
it is tempting to explore whether there are further approximations
to the outcome of this procedure that give satisfactory results
but require less computation.

Recall that $\pi_1, \ldots, \pi_{m+n}$ is the pooled collection
of placed reads and that $\tilde P = \frac{1}{m}\sum_{i \in I} \pi_i$ and
$\tilde Q = \frac{1}{n}\sum_{j \in J} \pi_j$, where $I$ is
a uniformly distributed random subset of $\{1, \ldots, m+n\}$
and $J$ is its complement.
Write
\[
G_k(u) := \pi_k(\tau(u)) \ \hbox{for any } u \in T, \ 1 \le k \le m+n,
\]
where we recall that $\tau(u)$ is the tree below $u$ relative to
the root $\rho$.  Define a $T$-indexed stochastic process
$X = (X(u))_{u \in T}$ by
\[
\begin{split}
X(u) & := \tilde P(\tau(u)) - \tilde Q(\tau(u)) \\
& = \frac{1}{m} \sum_{i \in I} G_i(u) - \frac{1}{n} \sum_{j \in J} G_j(u). \\
\end{split}
\]
Then,
\[
\kr_p(\tilde P, \tilde Q)
=
\left[\int_T \left| X(u) \right|^p \, \lambda(du)\right]^{\frac{1}{p} \wedge 1}.
\]

If $H_k$, $1 \le k \le m+n$, is the indicator random variable
for the event $\{k \in I\}$, then
\[
X(u) = \sum_{k=1}^{m+n} \left[\left(\frac{1}{m} + \frac{1}{n}\right) H_k - \frac{1}{n}\right] G_k(u).
\]
Writing $\mathbb{E}$, $\mathbb{V}$, and $\mathbb{C}$ for expectation,
variance, and covariance, we have
\[
\mathbb{E}[H_i] = \frac{m}{m+n},
\]
\[
\mathbb{V}[H_i] = \frac{m}{m+n} \frac{n}{m+n},
\]
and
\[
\mathbb{C}[H_i, H_j] = -\frac{1}{m+n-1} \frac{m}{m+n} \frac{n}{m+n}, \quad i \ne j.
\]
It follows that
\[
\mathbb{E}[X(u)] = 0
\]
and
\[
\begin{split}
& \mathbb{C}[X(u), X(v)] \\
& \quad = \frac{1}{mn}
\left( \sum_i G_i(u) G_i(v) - \frac{1}{m+n-1} \sum_{i \ne j} G_i(u) G_j(v) \right) \\
& \quad \approx \frac{1}{mn}
\left[ \sum_i G_i(u) G_i(v) - \frac{1}{m+n} \sum_{i,j} G_i(u) G_j(v) \right] \\
& \quad = \frac{1}{mn}
\left[ \sum_i G_i(u) G_i(v) - (m+n) \left(\frac{1}{m+n} \sum_i G_i(u)\right) \left(\frac{1}{m+n} \sum_j G_j(v)\right) \right] \\
& \quad = \frac{1}{mn}
\left[ \sum_i \left( G_i(u) - \bar G(u)\right) \left(G_i(v) - \bar G(v)\right) \right] \\
& \quad =: \Gamma(u,v) \\
\end{split}
\]
when $m+n$ is large, where $\bar G(u) := \frac{1}{m+n} \sum_k G_k(u)$.

\begin{remark}
\label{rem:point_placements}
In the case of point placements, with the probability distribution
$\pi_k$ being the point mass at $w_k \in T$ for $1 \le k \le m+n$, then
\[
\begin{split}
\Gamma(u,v)
=
\frac{1}{mn}
&
\left[
\sum_k
\#\{k : u \in [\rho, w_k], \, v \in [\rho, w_k]\} \right. \\
&
\qquad
\left.
-
\frac{1}{m+n} \#\{k : u \in [\rho, w_k]\} \#\{k : v \in [\rho, w_k]\}
\right]. \\
\end{split}
\]
\end{remark}

By a standard central limit theorem for exchangeable random variables
(see, e.g., Theorem 16.23 of Kallenberg, 2001), \nocite{kallenberg97}
the process $X$
is approximately Gaussian with covariance kernel $\Gamma$ when $m+n$ is large.
A straightforward calculation shows that
we may construct a Gaussian process $\xi$ with covariance kernel $\Gamma$
by taking independent standard Gaussian random variables
$\eta_1, \ldots, \eta_{m+n}$ and setting
\[
\xi(u) =
\frac{1}{\sqrt{mn}}
\left[
\sum_i G_i(u) \eta_i - \frac{1}{m+n} \left(\sum_i G_i(u)\right) \left( \sum_i \eta_i \right)
\right].
\]

It follows that the distribution of $\kr_p(\tilde P, \tilde Q)$ is approximately
that of the random variable
\begin{equation}
  \label{eq:intXi}
  \left[ \int_T |\xi(u)|^p \, \lambda(du) \right]^{\frac{1}{p} \wedge 1}.
\end{equation}
One can repeatedly sample the normal random variates $\eta_i$ and numerically integrate \eqref{eq:intXi} to approximate the distribution of this integral.
In the example application of Section~\ref{sec:application},
this provides a reasonable though not perfect approximation (Figure~\ref{fig:shuffnorm}).

There is an even simpler approach for the case $p=2$.
Let $\mu_k^2$, $k=1,2, \ldots$, and $\psi_k$, $k=1,2, \ldots$,
be the positive eigenvalues and corresponding normalized eigenfunctions
of the non-negative definite, self-adjoint, compact operator on
$L^2(\lambda)$ that maps the function $f$ to the function
$\int_T \Gamma(\cdot, v) f(v) \, \lambda(dv)$.  The
functions $\mu_k \psi_k$, $k=1,2, \ldots$, form an
orthonormal basis for the reproducing kernel Hilbert space
associated with $\Gamma$ and the Gaussian process $\xi$
has the Karhunen-Lo\`eve expansion
\[
\xi(u) = \sum_k \mu_k \psi_k(u) \eta_k,
\]
where $\eta_k$, $k=1,2, \ldots$, are independent
standard Gaussian random variables
-- see \citep{MR515431}
for a review of the theory of reproducing kernel Hilbert spaces
and the Karhunen-Lo\`eve expansion.

Therefore,
\[
\int_T |\xi(u)|^2 \, \lambda(du)
=
\sum_k \mu_k^2 \eta_k^2,
\]
and the distribution of $\kr_2^2(\tilde P, \tilde Q)$
is approximately that of a certain positive linear combination
of independent $\chi_1^2$ random variables.

The eigenvalues of the operator associated with
$\Gamma$ can be found by calculating the eigenvalues of a related matrix
as follows.
Define an $(m+n) \times (m+n)$ non-negative definite,
self-adjoint matrix $M$ given by
\[
M_{ij} :=
\frac{1}{mn}
\int_T \left(G_i(u) - \bar G(u)\right) \left(G_j(u) - \bar G(u)\right) \, \lambda(du).
\]

Note that if we have point placements at locations $w_k \in T$
for $1 \le k \le m+n$ as in
Remark~\ref{rem:point_placements}, then
\[
M
= \frac{1}{mn}
\left(I - \frac{1}{m+n} \mathbf{1} \mathbf{1}^\top \right)
N
\left(I - \frac{1}{m+n} \mathbf{1} \mathbf{1}^\top \right),
\]
where $I$ is the identity matrix, $\mathbf{1}$ is the vector which has $1$ for
every entry, and the matrix $N$ has $(i,j)$ entry given by the
distance from the root to the ``most recent common ancestor'' of $w_i$
and $w_j$.
% coding note: this N is the point placement version of mtilde.

Suppose that  $x$ is an eigenvector
of $M$ for the positive eigenvalue $\nu^2$. Set
\begin{equation}
  \label{eq:gammaEfun}
  \psi(u) := \sum_j  \left(G_j(u) - \bar G(u)\right) x_j.
\end{equation}
Observe that
\[
\begin{split}
& \int_T \Gamma(u,v) \psi(v) \, \lambda(dv) \\
& \quad =
\frac{1}{mn}
\int_T
\left[ \sum_i \left( G_i(u) - \bar G(u)\right) \left(G_i(v) - \bar G(v)\right) \right]
\sum_j  \left(G_j(v) - \bar G(v)\right) x_j
\, \lambda(dv) \\
& \quad =
\sum_i \left( G_i(u) - \bar G(u)\right)
\sum_j M_{ij} x_j \\
& \quad =
\sum_i \left( G_i(u) - \bar G(u)\right)
\nu^2 x_i \\
& \quad =
\nu^2 \psi(u), \\
\end{split}
\]
and so $\psi$ is an (unnormalized) eigenfunction of the operator on $L^2(\lambda)$
defined by the covariance kernel $\Gamma$ with eigenvalue $\nu^2$.

Conversely, suppose that $\mu^2$ is an eigenvalue of the operator with
eigenfunction $\phi$.  Set
\[
x_j := \int_T \left(G_j(v) - \bar G(v)\right) \phi(v) \, \lambda(dv).
\]
Then,
\[
\begin{split}
&
\sum_j M_{ij} x_j \\
& \quad =
\sum_j
\frac{1}{mn}
\int_T \left(G_i(u) - \bar G(u)\right) \left(G_j(u) - \bar G(u)\right) \, \lambda(du) \\
& \qquad \times
\int_T \left(G_j(v) - \bar G(v)\right) \phi(v) \, \lambda(dv) \\
& \quad =
\int_T \left(G_i(u) - \bar G(u)\right) \\
& \qquad \times \left[
\int_T
\frac{1}{mn}
\sum_j \left(G_j(u) - \bar G(u)\right)
\left(G_j(v) - \bar G(v)\right) \phi(v) \, \lambda(dv)
\right]
\, \lambda(du) \\
& \quad =
\int_T \left(G_i(u) - \bar G(u)\right)
\left[
\int_T
\Gamma(u,v) \phi(v) \, \lambda(dv)
\right]
\, \lambda(du) \\
& \quad =
\int_T \left(G_i(u) - \bar G(u)\right)
\mu^2 \phi(u) \, \lambda(du) \\
& \quad =
\mu^2 x_i, \\
\end{split}
\]
so that $\mu^2$ is an eigenvalue of $M$ with (unnormalized) eigenvector of $x$.

It follows that the positive eigenvalues of the operator
associated with $\Gamma$ coincide  with those of the matrix $M$
and have the same multiplicities.
%This is essentially just the
%linear algebra fact that if $A$ is a rectangular real matrix, then $A A^T$
%and $A^T A$ have the same non-zero eigenvalues with the same multiplicities,
%and if $x$ is an eigenvector of $A A^T$ with non-zero eigenvalue $\rho$, then
%$A^T x$ is an eigenvector of $A^T A$ with eigenvalue $\rho$.

However, we don't actually need to compute the eigenvalues of $M$
to implement this approximation.  Because $M$ is orthogonally equivalent
to a diagonal matrix with the eigenvalues of $M$ on the diagonal, we have
from the invariance under orthogonal transformations of the distribution
of the random vector $\eta := (\eta_1, \ldots, \eta_{m+n})^\top$ that
$\sum_k \mu_k^2 \eta_k^2$ has the same distribution as $\eta^\top M \eta$.
Thus, the distribution of the random variable $\kr_2^2(\tilde P, \tilde Q)$
is approximately that of $\sum_{ij} M_{ij} \eta_i \eta_j$.

One might hope to go even further in the $p=2$ case and obtain
an analytic approximation for the distribution $\sum_k \mu_k^2 \eta_k^2$ or
a useful upper bound for its right tail.
It is shown in \citep{MR548094} that if we order the positive eigenvalues so that
$\mu_1^2 \ge \mu_2^2 \ge \ldots$ and assume that $\mu_1^2 > \mu_2^2$, then
\[
\mathbb{P}\left\{\sum_k \mu_k^2 \eta_k^2 \ge r \right\}
\sim
\sqrt{\frac{2}{\pi}} \, \mu_1
\prod_{k>1} \left(1 - \frac{\mu_k^2}{\mu_1^2}\right)^{-\frac{1}{2}}
r^{-\frac{1}{2}} \exp\left(-\frac{r}{2 \mu_1^2}\right),
\]
in the sense that the ratio of the two sides converges to one as $r \to \infty$.
It is not clear what the rate of convergence is in this result and it appears
to require a detailed knowledge of the spectrum of the matrix $M$ to apply it.

Gaussian concentration inequalities such as Borell's inequality
(see, for example, Section 4.3 of \citep{MR1642391}) give bounds
on the right tail that only require a knowledge of
$\mathbb{E}[(\sum_k \mu_k^2 \eta_k^2)^{\frac{1}{2}}]$ and
$\mu_1^2$, but these bounds are far too conservative for the
example in Section~\ref{sec:application}.

There is a substantial literature
on various series expansions of densities of positive linear combinations of independent
$\chi_1^2$ random variables.   Some representative papers are
\citep{MR0032151, MR0067417, MR0067416, MR0137189, MR0211510, MR0403020}.
However, it seems that applying such results would also require a detailed
knowledge of the spectrum of the matrix $M$ as well as a certain amount
of additional computation to obtain the coefficients in the expansion and then
to integrate the resulting densities, and this may not be warranted given the relative
ease with which it is possible to repeatedly simulate the random variable
$\eta^\top M \eta$.

Even though these more sophisticated ways of using the Gaussian approximation
may not provide tight bounds, the process of repeatedly sampling normal random
variates $\eta_i$ and numerically integrating the resulting Gaussian
approximation \eqref{eq:intXi} does provide a useful way of approximating the
distribution obtained by shuffling. This approximation is significantly faster
to compute for larger collections of placements. For example, we considered a
reference tree with 652 leaves and 5 samples with sizes varying from 3372 to
15633 placements. For each of the 10 pairs of samples, we approximated the
distribution of the $Z_1$ distance under the null hypothesis of no difference
by both creating ``pseudo-samples'' via random assignment of reads to each
member of the pair (``shuffling'') and by simulating the Gaussian process
functional with a distribution that approximates that of the $Z_1$ distance
between two such random pseudo-samples. We used 1000 Monte Carlo steps for both
approaches. The (shuffle, Gaussian) run-times in seconds ranged from (494.1,
36.8) to (36.1, 2.2); in general, the Gaussian procedure ran an order of
magnitude faster than the shuffle procedure.

\subsection{Interpretation of p-values}

Although the above-described
permutation procedure is commonly used
to assess the statistical significance of an observed distance,
we discuss in this section
how its interpretation is not completely straightforward.

In terms of the classical Neyman-Pearson framework for hypothesis testing, we are computing a p-value for
the null hypothesis that an observed subdivision of
a set of $m+n$ objects into two groups of size $m$ and $n$
looks like a uniformly distributed random subdivision against
the complementary alternative hypothesis.
For many purposes, this turns out to be a reasonable proxy for the imperfectly-defined notion that the two groups are ``the same''
rather than ``different.''

However, a rejection of the null hypothesis may not have
the interpretation that is often sought
in the microbial context -- namely, that the
two collections of reads
represent communities that are different in biologically
relevant ways.
For example, assume that $m=n=NK$ for integers $N$ and $K$.  Suppose that the placements in each sample
are obtained by independently laying down
$N$ points uniformly (that is, according to the normalized version of the measure $\lambda$) and then putting $K$ placements at
each of those points.
The stochastic mechanism generating the two samples is identical
and they are certainly not different in any interesting way, but
if $K$ is large relative to $N$ the resulting collections of placements
will exhibit a substantial ``clustering'' that will be
less pronounced in the random pseudo-samples,
and the
randomization procedure will tend to produce a
``significant'' p-value for the
observed KR distance if the clustering is not taken into account.

These considerations motivate consideration of randomization tests performed on data which is ``clustered'' on an organismal level.
Clustering reads by organism is a difficult task and
an active research topic \citep{WhiteEaAlignmentClustering10}.
A thoroughgoing exploration of the effect of different
clustering techniques is beyond the scope of this paper, but we examine the impact of some simple approaches in the next section.

\section{Example application}
\label{sec:application}

\FORarxiv{\FIGcontrolTree}
\FORarxiv{\FIGdmspTree}

To demonstrate the use of the $Z_p$ metric in an example application, we investigated variation in expression levels for the \psbA\ gene for an experiment in the Sargasso Sea \citep{villaCostaEaDMSP10}.
Metatranscriptomic data was downloaded from the CAMERA website (http://camera.calit2.net/), and a \psbA\ alignment was supplied by Robin Kodner.
Searching and alignment was performed using HMMER \citep{eddyProfileHMMs98}, a reference tree was inferred using RAxML \citep{stamatakisEaRAxML06}, and phylogenetic placement was performed using pplacer \citep{matsenEaPplacer10}.
The calculations presented here were performed using the ``guppy'' binary available as part of the pplacer suite of programs (\hbox{http://matsen.fhcrc.org/pplacer}).

\FORarxiv{\FIGshuffnorm}

Visual inspection of the trees fattened by number of placements showed the same overall pattern with some minor differences (Figure~\ref{fig:controlTree} and \ref{fig:dmspTree}).
However, application of the KR metric revealed a significant difference between the two samples.
The value of $Z_1$ for this example (using spread placements and normalizing by total tree length) was 0.006601. This is far out on the tail of the distribution (Figure~\ref{fig:shuffnorm}), and is in fact larger than any of the 1000 replicates generated via shuffling or the Gaussian-based approximation.

\FORarxiv{\FIGbary}

Such a low p-value prompts the question of whether the center of mass of the two distributions is radically different in the two samples (see Section~\ref{sec:barycenter}).
In this case, the answer is no, as the two barycenters are quite close together (Figure~\ref{fig:bary}; see Section~\ref{sec:barycenter}).

It was not intuitively obvious to us how varying $p$ would affect the distribution
of the $Z_p$ distance under the null hypothesis of no clustering.
To investigate this question, we plotted the observed distance
along with boxplots of the null distribution for a collection of different $p$ (Figure~\ref{fig:box}). It is apparent that there is a consistent
conclusion over a wide range of values of $p$.

\FORarxiv{\FIGbox}

One can also visualize the difference between the two samples by drawing the reference tree with branch thicknesses that represent the minimal amount of ``mass'' that flows through that branch
in the optimal transport of mass implicit
in the computation of $Z_1(P,Q)$
and with branch shadings that indicate the sign of the movement
(Figure~\ref{fig:heat}).

\FORarxiv{\FIGheat}

Next we illustrate the impact of simple clustering on randomization tests for KR.
The clustering for these tests will be done by rounding placement locations using two parameters: the mass cutoff $C$ and the number of significant figures $S$ as follows.
Placement locations with low probability mass for a given read are likely to be error-prone \citep{matsenEaPplacer10}, thus the first step is to through away those locations associated with posterior probability or ``likelihood weight ratio'' below $C$.
The second step is to round the placement attachment location and pendant branch length by multiplying them by $10^S$ and rounding to the nearest integer.
The reads whose placements are identical after this rounding process are then said to cluster together.
We will call the number of reads in a given cluster the ``multiplicity'' of the cluster.

After clustering, various choices can be made about how to scale the mass distribution according to multiplicity.
Again, each cluster has some multiplicity and a distribution of mass across the tree according to likelihood weight.
One option (which we call straight multiplicity) is to multiply the mass distribution by the multiplicity.
Alternatively, one might forget about multiplicity by distributing a unit of mass for each cluster irrespective of multiplicity.
Or one might do something intermediate by multiplying by a transformed version of multiplicity; in this case we transform by the hyperbolic arcsine, $\asinh$.

\FORarxiv{\TABrounding}

We calculated distances and p-values for several clustering parameters and multiplicity uses (Table~\ref{tab:rounding}).
To randomize a clustered collection of reads, we reshuffled the labels on the clusters, maintaining the groupings of the reads within the clusters; thus, all the placements in a given cluster were
assigned to the same pseudo-sample.
The distances do not change very much under different collections of clustering parameters, as there is little redistribution of mass.
On the other hand, the p-values are different, because under our randomization strategy mass is relabeled on a cluster-by-cluster level.
The different choices represented in this table represent different perspectives on what the multiplicities mean.
The ``strict'' multiplicity-based p-value corresponds to interpreting the multiplicity with which reads appear as meaningful, the unit cluster p-value corresponds to interpreting the multiplicities as noise, and the $\asinh$-transformed multiplicity sits somewhere in between.
The p-value with no clustering (as above, $Z_1 = 0.006601$, with a p-value of 0) corresponds to interpreting reads as being sampled one at a time from a distribution.

The choice of how to use multiplicity information depends on the biological setting.
There is no doubt that increased organism abundance increases the likelihood of sampling a read from that organism, however the relationship is almost certainly nonlinear and dependent on species and experimental setup \citep{morganEaInVitroMetagenome10}.
How multiplicities are interpreted and treated in a specific instance
is thus a decision that is best left to the
researcher using his/her knowledge of the environment
being studied and the details of the experimental procedure.

\section{Discussion}
\label{sec:discussion}

\subsection{Other approaches}

\subsubsection{Operational Taxonomic Units (OTUs)}
The methods described in this paper are complementary to comparative methods based on ``operational taxonomic units'' (OTUs).
OTUs are groups of reads which are assumed to represent the reads from a single species, and are typically heuristically defined using a fixed percentage sequence similarity cutoff.
A comparative analysis then proceeds by comparing the frequency of various OTUs in the different samples.
There has been some contention about whether OTU-based methods or phylogenetic based methods are superior-- e.g. \citep{SchlossEvaluating08} and \citep{LozuponeEaUniFracEffective10}-- but most studies use a combination of both, and the major software packages implement both.
A recent comparative study for distances on OTU abundances can be found in a paper by \citet{KuczynskiEaMicrobialResemblance10}.

\subsubsection{Other phylogenetic approaches}
There are a number of ways to compare microbial samples in a phylogenetic context besides the method presented here.
One popular means of comparing samples is the ``parsimony test,'' by which the most parsimonious assignment of internal nodes of the phylogenetic tree to communities is found; the resulting parsimony score is interpreted as a measure of difference between communities \citep{slatkinMaddisonTest89, schlossHandelsmanTreeClimber06}.
Another interesting approach is to consider a ``generalized principal components analysis'' whereby the tree structure is incorporated into the process of finding principal components of the species abundances \citep{bikEaStomach06, purdomAnalyzingDataWithGraphs08}.
The Kantorovich-Rubinstein metric complements these methods by providing a means of comparing samples that leverages established statistical methodology, that takes into account uncertainty in read location, and can be visualized directly on the tree.

There are other metrics that could be used to compare probability distributions on a phylogenetic tree.
The metric on probability distributions that is most familiar to statisticians other
than the total variation distance is probably the Prohorov metric and so they
may feel more comfortable using it rather than the KR metric.  However, the
Prohorov metric is defined in terms of an optimization that does not
appear to have a closed form solution on a tree and, in any case, for
a compact metric space there
are results that bound the Prohorov metric above
and below by functions of the KR metric
(see Problem 3.11.2 of Ethier and Kurtz; 1986 \nocite{MR838085})
so the two metrics incorporate very similar information
about the differences between a pair of distributions.

\subsection{The barycenter of a probability distribution on a phylogenetic tree}
\label{sec:barycenter}

It can be useful to compare probability distributions on a metric space by calculating a suitably
defined ``center of mass'' that provides a single point summary for each distribution.
Recall the standard fact that if $P$ is a probability distribution
on a Euclidean space such that $\int |y-x|^2 \, P(dy)$ is finite for some (and hence all) $x$,
then the function $x \mapsto \int |y-x|^2 \, P(dy)$ has a unique minimum at $x_0 = \int y \, P(dy)$.
A probability distribution $P$ on an arbitrary metric space $(S,r)$
has a ``center of mass'' or {\em barycenter} at $x_0$ if $\int r(x,y)^2 \, P(dy)$ is finite for some (and hence all $x$)
and the function $x \mapsto \int r(x,y)^2 \, P(dy)$ has a unique minimum at $x_0$.
In terms of the concepts introduced above, the
barycenter is the point $x$ that minimizes the $Z_2$ distance between the point mass $\delta_x$ and $P$.

Barycenters need not exist for general metric spaces.
However, it is well-known that barycenters do exist for
probability distributions on {\em Hadamard spaces}.
A Hadamard space is a simply
connected complete metric space in which there is a suitable notion of the length of a path in the space,
the distance between two points is the infimum of the lengths of the paths joining the points,  and the space
has nonpositive curvature in an appropriate sense -- see \citep{MR1835418}.  Equivalently,
a Hadamard space is a complete $\mathrm{CAT}(0)$ space in the sense of \citep{MR1744486}.

It is a straightforward
exercise to check that a tree is a Hadamard space -- see Example II.1.15(4) of \citep{MR1744486}
and note the remark after Definition II.1.1 of \citep{MR1744486} that a Hadamard space
is the same thing as a complete $\mathrm{CAT}(0)$ space.
Note that $\mathrm{CAT}(0)$ spaces have already made an appearance in phylogenetics in the description of spaces of phylogenetic trees \citep{billeraEaGeometry01}.

The existence of barycenters
on the tree $(T,d)$ may also be established directly as follows.
As a continuous function on a compact metric space, the function $f:T \to \mathbb{R}_+$ defined by $f(x) := \int_T d(x,y)^2 \, P(dy)$ achieves its infimum.
Suppose that the infimum is achieved at two points $x'$ and $x''$. Define a function $\gamma:[0,d(x',x'')] \to [x',x'']$, where $[x',x''] \subseteq T$ is the arc between $x'$ and $x''$, by the requirement that $\gamma(t)$ is the unique point in $[x',x'']$ that is distance $t$ from $x'$.  It is straightforward to check that the composition $f \circ \gamma $ is {\em strongly convex}; that is,
\[
(f\circ \gamma) (\alpha r + (1 - \alpha) s) < \alpha (f\circ \gamma)(r) + (1 - \alpha) (f\circ \gamma)(s)
\]
for $0 < \alpha < 1$ and $r,s \in [0, d(x',x'')]$.
In particular, $f(\gamma(d(x',x'')/2)) = (f \circ \gamma)(d(x',x'')/2) < (f(x') + f(x''))/2$, contradicting the definitions of $x'$ and $x''$.
Thus, a probability distribution on a tree possesses a barycenter in the above sense.

We next consider how to compute the barycenter of a probability distribution $P$ on the tree $(T,d)$.
For each point $u \in T$ there is the associated set of directions in which it is possible to proceed
 when leaving $u$.  There is one direction for every connected component of $T \setminus \{u\}$.
 Thus, there is just one direction associated with a leaf, two directions associated with
 a point in the interior of a branch, and $k$ associated with a vertex of degree $k$.
Given a point $u$ and a direction $\delta$, write $T(u,\delta)$ for the subset of $T$ consisting
of points $v \ne u$ such that the unique path connecting $u$ and $v$ departs $u$ in the direction $\delta$,
set
\[
D(u,\delta) := -\int_{T(u,\delta)} d(u,y) \, P(dy) +  \int_{T \setminus T(u,\delta)} d(u,y) \, P(dy),
\]
and note that
\[
\lim_v \, \frac{1}{d(u,v)} \left[ \int_T d(v,y)^2 \, P(dy) - \int_T d(u,y)^2 \, P(dy) \right]
=
2 D(u,\delta),
\]
where the limit is taken over $v \to u$, $v \in T(u,\delta)$.
Note that if $u$ is in the
interior of a branch $[a,b]$ and $b$ is in the direction $\delta$ from $u$, $u$ is in
the direction $\alpha$ from $a$, and $u$ is in
the direction $\beta$ from $b$, then
\[
\begin{split}
D(u,\delta)
& =
-\int_{T \setminus T(b,\beta)} d(u,y) \, P(dy) -  \int_{(u,b)} d(u,y) \, P(dy) \\
& \quad + \int_{T \setminus T(a,\alpha)}   d(u,y) \, P(dy) +  \int_{(a,u)} d(u,y)  \, P(dy) \\
& =
-\int_{T \setminus T(b,\beta)} d(a,y) \, P(dy) + d(a,u) P(T \setminus T(b,\beta)) \\
& \quad - \int_{(u,b)} d(a,y) \, P(dy) + d(a,u) P((u,b)) \\
& \quad + \int_{T \setminus T(a,\alpha)}   d(a,y) \, P(dy) + d(a,u) P(T \setminus T(a,\alpha)) \\
& \quad + d(a,u) P((a,u)) - \int_{(a,u)} d(a,y)  \, P(dy) \\
& =
D(a,\alpha) + d(a,u). \\
\end{split}
\]

If for some vertex $u$ of the reference tree, $D(u,\delta)$ is greater than 0 for all directions $\delta$ associated
with $u$, then $u$ is the barycenter (this case includes the trivial one in which $u$ is a leaf
and all the mass of $P$ is concentrated on $u$).
If there is no such vertex, then there must be a unique pair of neighboring vertices $a$ and $b$ such that
$D(a,\alpha) < 0$ and $D(b,\beta) < 0$, where
$\alpha$ is the direction from $a$ pointing towards $b$ and
$\beta$ is the direction from $b$ pointing towards $a$.
In that case, the barycenter must lie on the branch between $a$ and $b$, and it follows
from the calculations above that the
barycenter is the point $u \in (a,b)$ such that $d(a,u) = -D(a,\alpha)$.

\subsection{$Z_2^2(P,Q)$ and ANOVA}
\label{sec:ANOVA}

In this section we demonstrate how $\kr_2^2(P,Q)$ can be interpreted as a difference between the pooled average of pairwise distances and the average for each sample individually.

As above, let $\pi_1, \ldots, \pi_m$ (resp. $\pi_{m+1}, \ldots, \pi_{m+n}$) be the placements
in the first (resp. second) sample, so that each $\pi_k$ is a probability distribution on the
tree $T$, $P = \frac{1}{m} \sum_{i=1}^m \pi_i$, and $Q = \frac{1}{n} \sum_{j=m+1}^{m+n} \pi_j$.
Set
\[
R := \frac{m}{m+n} P + \frac{n}{m+n} Q = \frac{1}{m+n} \sum_k \pi_k.
\]
Recall the $T$-valued random variables $X',X'',Y',Y''$
that appeared in \eqref{eq:Z2_equivalent}.  If $I',I''$
are $\{0,1\}$-valued random variables with
$\mathbb{P}\{I'=1\} = \mathbb{P}\{I''=1\} = \frac{m}{m+n}$
and $X',X'',Y',Y'', I', I''$ are independent, then defining
$Z',Z''$ by
$Z' =  X'$ on the event $\{I'=1\}$ (resp. $Z'' =  X''$ on the event $\{I''=1\}$)
and
$Z' =  Y'$ on the event $\{I'=0\}$ (resp. $Z'' =  Y''$ on the event $\{I''=0\}$)
gives two $T$-valued random variables with common distribution $R$.
%\[
%\begin{split}
%2 mn \int_T \int_T d(v,w) \, P(dv) \, Q(dw)
%& = (m+n)^2 \int_T \int_T d(v,w) \, R(dv) \, R(dw) \\
%& \quad - m^2 \int_T \int_T d(v,w) \, P(dv) \, P(dw) \\
%& \quad - n^2 \int_T \int_T d(v,w) \, Q(dv) \, Q(dw) \\
%\end{split}
%\]
%and

It follows readily from \eqref{eq:Z2_equivalent} that
\[
\begin{split}
& \kr_2^2(P,Q) \\
& \quad =
\frac{1}{2}
\frac{(m+n)^2}{mn}
\biggl[
\mathbb{E}\left[ d(Z',Z'') \right]
-
\left\{
\frac{m}{m+n} \mathbb{E}\left[ d(X',X'') \right]
+
\frac{n}{m+n} \mathbb{E}\left[ d(Y',Y'') \right]
\right\}
\biggr]\\
%& =
%\frac{1}{2}
%\biggl[
%\frac{(m+n)^2}{mn} \int_T \int_T d(v,w) \, R(dv) \, R(dw) \\
%& \quad - \frac{(mn + m^2)}{mn} \int_T \int_T d(v,w) \, P(dv) \, P(dw) \\
%& \quad - \frac{(mn + n^2)}{mn} \int_T \int_T d(v,w) \, Q(dv) \, Q(dw)
%\biggr]\\
& \quad =
\frac{1}{2}
\frac{(m+n)^2}{mn}
\biggl[
\int_T \int_T d(v,w) \, R(dv) \, R(dw) \\
& \qquad -
\biggl\{
\frac{m}{m+n} \int_T \int_T d(v,w) \, P(dv) \, P(dw)
+
\frac{n}{m+n} \int_T \int_T d(v,w) \, Q(dv) \, Q(dw)
\biggr\}
\biggr].\\
\end{split}
\]
Thus, $\kr_2^2(P,Q)$ gives an indication of the ``variability'' present
in the pooled collection $\pi_k$, $1 \le k \le m+n$, that is
over and above the ``variability'' in the two collections $\pi_i$, $1 \le i \le m$,
and $\pi_j$, $m+1 \le j \le m+n$.

\section{Conclusion}

As sequencing becomes faster and less expensive, it will become increasingly common to have a collection of large data sets for a given gene.
Phylogenetic placement can furnish an evolutionary context for query sequences, resulting in each data set being represented as a probability distribution on a phylogenetic tree.
The Kantorovich-Rubinstein metric is a natural means to compare those probability distributions.
In this paper we showed that the weighted UniFrac metric is the phylogenetic Kantorovich-Rubinstein metric for point placements.
We explored Zolotarev-type generalizations of the KR metric, showed how to approximate the limiting distribution and made connections with the analysis of variance.

The phylogenetic KR metric and its generalizations can be used any time one wants to compare two probability distributions on a tree.
However, our software implementation is designed with metagenomic and metatranscriptomic investigations in mind; for this reason it is tightly integrated with the phylogenetic placement software pplacer \citep{matsenEaPplacer10}.
With more than two samples,
principal components analysis and hierarchical clustering can be applied to the pairwise distances
to cluster environments based on the KR distances as has been done with UniFrac \citep{lozuponeKnightUniFrac05, lozuponeEaGutCarbohydrate08, hamadyEaFastUniFrac09}.
We have recently developed versions of these techniques which leverage the special structure of this data \citep{edgeSquash}.

% Further work will also combine information from a large collection of genes to deliver a comparative method on the level of metagenomes.
% In a large scale metagenomic analysis with several genes, the KR and related metrics could be used to scan for ``discriminating'' genes, i.e. genes whose distribution differs between environments.
Another potential future extension not explored here is to partition the tree into subsets in a principal components fashion for a single data set.
Recall that \eqref{eq:gammaEfun} gives a formula for the eigenfunctions of the covariance kernel $\Gamma$ given the eigenvectors of $M$.
For any $k$, one could partition the tree into subsets based on the sign of the product of the first $k$ eigenfunctions, which would be analogous to partitioning Euclidean space by the hyperplanes associated with the first $k$ eigenvectors in traditional principal components analysis.

Future methods will also need to take details of the DNA extraction procedure into account.
Recent work shows that current lab methodology is unable to recover absolute mixture proportions due to differential ease of DNA extraction between organisms \citep{morganEaInVitroMetagenome10}.
However, relative abundance between samples for a given organism with a fixed laboratory protocol potentially can be measured, assuming consistent DNA extraction protocols are used.
An important next step is to incorporate such organism-specific biases into the sort of analysis described here.

\FORsubmit{
\newpage
\section{Tables and figures}

\TABrounding
\FIGcontrolTree
\FIGdmspTree
\FIGshuffnorm
\FIGbary
\FIGbox
\FIGheat
}

\clearpage

\appendix

\section*{Acknowledgements}
The authors are grateful to Robin Kodner for her \psbA\ alignment,
the Armbrust lab for advice and for the use of their computing
cluster, Mary Ann Moran and her lab for allowing us to use her
metagenomic sample from the DMSP experiment, David Donoho for an
interesting suggestion, Steve Kembel for helpful conversations, and
Aaron Gallagher for programming support.

The manuscript was greatly improved by suggestions from one of the editors,
an associate editor, and two anonymous reviewers.

\bibliographystyle{Chicago}
\bibliography{phylo_kr}

\end{document}